\documentclass[usenatbib]{mn2e} 
\usepackage{graphicx,fixltx2e}

\def\rpd{\hbox{rad\,d$^{-1}$}}

\def\chisq{\hbox{$\chi^2$}}
\def\chisqr{\hbox{$\chi^2_{\rm r}$}}
\def\msun{\hbox{${\rm M}_{\odot}$}}
\def\mspy{\hbox{${\rm M}_{\odot}$\,yr$^{-1}$}}
\def\rsun{\hbox{${\rm R}_{\odot}$}}
\def\lsun{\hbox{${\rm L}_{\odot}$}}
\def\rcor{\hbox{$r_{\rm C}$}}
\def\rmag{\hbox{$r_{\rm A}$}}
\def\mstar{\hbox{$M_{\star}$}}
\def\rstar{\hbox{$R_{\star}$}}
\def\lstar{\hbox{$L_{\star}$}}
\def\teff{\hbox{$T_{\rm eff}$}}
\def\logg{\hbox{$\log g$}}
\def\sn{\hbox{S/N}}
\def\vrad{\hbox{$v_{\rm rad}$}}

\def\epspcs{\hbox{erg\,s$^{-1}$\,cm$^{-2}$}}
\def\kms{\hbox{km\,s$^{-1}$}}
\def\vsini{\hbox{$v \sin i$}}

\def\ptt{\hbox{$10^{-4} I_{\rm c}$}}

\def\arcsec{\hbox{$^{\prime\prime}$}}
\def\degr{\hbox{$^\circ$}}

\def\Mdot{\hbox{$\dot{M}$}}
\def\omeq{\hbox{$\Omega_{\rm eq}$}}
\def\dom{\hbox{$d\Omega$}}

\newcommand{\caii}{\hbox{Ca$\;${\sc ii}}}
\newcommand{\fei}{\hbox{Fe$\;${\sc i}}}

\newcommand{\hei}{\hbox{He$\;${\sc i}}}
\newcommand{\hal}{\hbox{H${\alpha}$}}
\newcommand{\hbe}{\hbox{H${\beta}$}}

\begin{document}

\title[Magnetic topology and differential rotation of V2247~Oph]{Complex magnetic topology 
and strong differential rotation on the low-mass T~Tauri star V2247~Oph\thanks{Based on observations 
obtained at the Canada-France-Hawaii Telescope, operated by the National Research Council of 
Canada, the Institut National des Sciences de l'Univers of the Centre National de la Recherche 
Scientifique of France and the University of Hawaii.  } }

\makeatletter

\def\newauthor{%
  \end{author@tabular}\par
  \begin{author@tabular}[t]{@{}l@{}}}
\makeatother
 
\author[J.-F.~Donati et al.]
{\vspace{1.7mm}
J.-F.~Donati$^1$\thanks{E-mail: 
donati@ast.obs-mip.fr (J-FD); 
mskelly@ast.obs-mip.fr (MBS); 
jerome.bouvier@obs.ujf-grenoble.fr (JB); 
mmj@st-andrews.ac.uk (MMJ); 
sg64@st-andrews.ac.uk (SGG); 
jmorin@ast.obs-mip.fr (JM); 
ghussain@eso.org (GAJH); 
catherine.dougados@obs.ujf-grenoble.fr (CD); 
francois.menard@obs.ujf-grenoble.fr (FM); 
y.unruh@imperial.ac.uk (YU)
}, 
M.B.~Skelly$^1$, J.~Bouvier$^2$, M.M.~Jardine$^3$, S.G.~Gregory$^{3,4}$, \\ 
\vspace{1.7mm}
{\hspace{-1.5mm}\LARGE\rm 
J.~Morin$^1$, G.A.J.~Hussain$^5$, C.~Dougados$^2$, F.~M\'enard$^2$, Y.~Unruh$^6$ } \\ 
$^1$ LATT--UMR 5572, CNRS \& Univ.\ de Toulouse, 14 Av.\ E.~Belin, F--31400 Toulouse, France \\
$^2$ LAOG--UMR 5571, CNRS \& Univ.\ J.~Fourier, 414 rue de la Piscine, F--38041 Grenoble, France \\ 
$^3$ School of Physics and Astronomy, Univ.\ of St~Andrews, St~Andrews, Scotland KY16 9SS, UK \\
$^4$ School of Physics, Univ.\ of Exeter, Stocker Road, Exeter EX4~4QL, UK \\ 
$^5$ ESO, Karl-Schwarzschild-Str.\ 2, D-85748 Garching, Germany \\ 
$^6$ Department of Physics, Imperial College London, London SW7 2AZ, UK 
}

\date{2009, MNRAS, submitted}
\maketitle
 
\begin{abstract}  

From observations collected with the ESPaDOnS spectropolarimeter at the 
Canada-France-Hawaii Telescope (CFHT), 
we report the detection of Zeeman signatures on the low-mass classical T~Tauri star 
(cTTS) V2247~Oph.  Profile distortions and circular polarisation signatures detected 
in photospheric lines can be interpreted as caused by cool spots and magnetic 
regions at the surface of the star.  The large-scale field is of moderate strength and highly 
complex;  moreover, both the spot distribution and the magnetic field show significant 
variability on a timescale of only one week, as a likely result of strong 
differential rotation.  Both properties make V2247~Oph very different from the 
(more massive) prototypical cTTS BP~Tau;  we speculate that this difference reflects 
the lower mass of V2247~Oph.  

During our observations, V2247~Oph was in a low-accretion state, with emission 
lines showing only weak levels of circular polarisation;  we nevertheless find that 
excess emission apparently concentrates in a mid-latitude region of strong radial 
field, suggesting that it is the footpoint of an accretion funnel.  

The weaker and more complex field that we report on V2247~Oph may share similarities 
with those of very-low-mass late-M dwarfs and potentially explain why low-mass cTTSs 
rotate on average faster than intermediate mass ones.  These surprising results need 
confirmation from new independent data sets on V2247~Oph and other similar low-mass 
cTTSs.  

\end{abstract}


\section{Introduction} 
\label{sec:int}

Whereas our understanding of most phases of stellar evolution made considerable 
progress throughout the twentieth century, stellar formation remained rather 
enigmatic and poorly constrained by observations until about two decades ago with 
the advent of the most powerful ground-based and space telescopes.  One of the 
major discoveries obtained then is that protostellar accretion discs are 
often associated with extremely powerful and highly collimated jets escaping the 
systems along their rotation axis \citep[e.g.,][]{Snell80}.  This finding has 
revolutionized the field of stellar formation;  in particular, it provided us 
with strong evidence that magnetic fields are playing an important role throughout 
stellar formation.  

Magnetic fields are expected to modify significantly the contraction of molecular 
clouds into protostellar cores and discs \citep[e.g.,][for a review]{Andre08}.  
They are also expected to influence the next formation stage, at least for stars 
with masses lower than about 2.5~\msun;  these protostars (called classical T~Tauri 
stars or cTTSs) apparently host magnetic fields strong enough to disrupt the central 
regions of their accretion discs, connect the protostars to the inner disc rim 
through discrete accretion funnels and possibly even slow down the rotation of 
protostars through the resulting star/disc magnetic torque \citep[e.g.,][for a 
review]{Bouvier07}.

While magnetic fields have been repeatedly reported over the last decade at the 
surface of a number of prototypical TTSs \citep[e.g.,][]{Johns99b, Johns07}, it 
is only recently that their large-scale topologies can be investigated into 
some details through phase-resolved spectropolarimetric observations 
\citep{Donati07, Donati08b, Hussain09}.  This new option offers a direct 
opportunity to pin down the role of magnetic fields in this crucial phase of the 
formation process.  In particular, it straightforwardly allows studying how magnetic 
topologies of cTTSs depend on mass and rotation rate, giving hints on the potential 
origin of such fields;  moreover, it provides material for quantitatively studying  
how protostars magnetically connect and interact with their accretion discs.  

Up to now, magnetic topologies of 4 cTTSs have been investigated 
with this method.  Among these 4, only BP~Tau ($\mstar\simeq0.7$~\msun) is 
fully convective and hosts a rather simple, dominantly dipolar, large-scale 
magnetic field, the 3 others being partly convective (as a result of their 
relatively higher mass, in excess of 1.3~\msun) and hosting a more complex magnetic 
topology.  In particular, this transition from simple to complex fields for stars 
on either sides of the full convection limit is reminiscent of the magnetic 
properties of low-mass main-sequence dwarfs \citep{Morin08b, Donati08d};  in 
addition to suggesting that magnetic fields of cTTSs are likely of dynamo origin, 
it also gives potential hints on why magnetospheric accretion and rotation properties 
of cTTSs are discrepant on both sides of the full convection limit (e.g., with higher 
mass cTTSs rotating more quickly in average).  

Extending this magnetic survey to a larger sample of cTTSs is needed to go 
further;  this is the exact purpose of the MaPP (Magnetic Protostars and Planets) 
program, in the framework of the international MagIcS research initiative 
(aimed at qualifying the magnetic properties of stars throughout the HR diagram).  
This survey has been allocated 690~hr of observing time on the 3.6~m 
Canada-France-Hawaii Telescope (CFHT) over 9 successive semesters (from 2008b to 2012b).  
The present paper concentrates on the low-mass\footnote{Throughout this paper, cTTSs with 
masses lower than 0.5~\msun, ranging from 0.5 to 1~\msun\ and larger than 1~\msun\ are 
respectively called low-mass, intermediate-mass and high-mass cTTSs. } 
cTTS V2247~Oph (spectral type M1) in the $\rho$~Oph 
star forming region (Lynds~1688 dark cloud);  
with a mass of only about 0.35~\msun\ (see Sec.~\ref{sec:v22}), 
it is less massive than all cTTSs yet magnetically imaged and thus appears as an ideal 
candidate to expand our sample towards cooler cTTSs.

\section{V2247~Oph = SR~12 = ROX~21 = HBC~263}
\label{sec:v22}

First classified as a TTS with weak emission lines (weak-line TTS or wTTS) from 
its moderate \hal\ emission \citep[equivalent width of order 0.4~nm, 
e.g.,][]{Bouvier92}, V2247~Oph has recently been reported to undergo transient 
episodes of stronger \hal\ emission \citep[equivalent width of up to 
1.76~nm][]{Littlefair04} that makes it qualify as a cTTS based on that criterion 
\citep[i.e., with an \hal\ emission larger than 0.9~nm at spectral type M1, 
e.g.,][]{Fang09}.  It therefore suggests that V2247~Oph is still surrounded by 
a (likely low-mass) accretion disc.  

The exact nature and evolutionary stage of this disc is however unclear.  Given 
that accretion is still on-going (although sporadically), we can assume that the 
inner disc contains at least gas.  However, V2247~Oph shows no significant infrared excess 
up to about 10~$\mu$m \citep{Gras05, Cieza07} indicating that the disc contains 
very little hot-dust.  ISOPHOT measurements indicate large excesses at 60 and 
100~$\mu$m \citep{Gras05} suggesting the presence of an outer disc;  however, 
given the large size of the aperture used (120\arcsec) and the IR-bright nebulosity 
located very close to V2247~Oph \citep[e.g.,][]{Padgett08}, this excess may potentially 
be unrelated to V2247~Oph itself and rather result (at least partly) from 
background/foreground contamination.  It thus awaits for a confirmation from 
higher spatial resolution data (e.g., Spitzer MIPS observations). 

V2247~Oph is also known to be a binary star with a companion located at about 
0.3\arcsec\ \citep{Simon87} from the primary star, i.e., at 42~AU assuming a distance 
of $139\pm6$~pc for V2247~Oph \citep[e.g.,][]{Mamajek08}.  
The presence of the companion likely implies that the disc features a gap at a few 
tens of AUs, with the inner disc being fed by a circumbinary structure (possibly that 
producing the far-IR excesses);  
this configuration may potentially generate lower-than-average and unsteady 
accretion rates like those observed on V2247~Oph.  
At most of the optical wavelengths we are working at in the present paper, the 
companion is outshone by the primary star by more than an order of magnitude and 
can therefore be neglected for our analysis.  

Fits to the spectral energy distribution \citep{Gras05} indicate that V2247~Oph has a 
photospheric temperature \teff\ of 3430~K, a radius \rstar\ and a luminosity \lstar\ of 
2.2~\rsun\ and 0.61~\lsun\ respectively (once corrected for a distance of 139~pc) and 
suffers a visual extinction of $A_{\rm V}\simeq0.45$.  This is in rough agreement with 
previous determinations from both photometric colors and spectroscopic indices 
\citep[e.g.,][]{Bouvier92}.  We thus assume here that $\teff=3500$~K (with a typical 
error of 150~K) and $\log\lstar/\lsun=-0.2$ (with a typical error of 0.1~dex).  
Fitting the evolutionary models of \citet{Siess00} to these parameters, we infer that 
V2247~Oph is a fully convective star with a mass of 0.36~\msun\ (possible values 
ranging from 0.29 to 0.45~\msun), a radius of $2.0\pm0.3$~\rsun\ and an age of about 
1~Myr.  Note that despite the large uncertainty on mass \citep[about 25\% and essentially 
reflecting the uncertainty on effective temperature,][]{Siess01}, we can safely conclude 
that V2247~Oph is truly a low-mass cTTS and in particular that it is significantly less massive 
than the intermediate-mass cTTS BP~Tau.  In the following, we assume that 
$\mstar=0.35$~\msun\ and $\rstar=2$~\rsun.  

From photometric modulation presumably caused by the presence of cool surface spots 
going in and out of view of Earth-based observers, V2247~Oph is reported to be 
rotating with a period of about 3.5~d \citep{Bouvier89, Schevchenko98}.  Long term 
photometric monitoring moreover reveals that this period 
is significantly changing with time \citep[from about 3.4 to 3.6~d,][]{Grankin08};  
it suggests that V2247~Oph \citep[as many other cool active stars, e.g.,][]{Hall91} 
is experiencing photospheric shearing due to latitudinal differential rotation, 
which generates different modulation periods for spots located at different latitudes.  

With $R_{\rm c}-I_{\rm c}=1.3$ \citep{Bouvier92}, V2247~Oph is located at a critical position in an 
HR diagram of young stars \citep{Lamm05}.  For stars bluer than this value (i.e.\ 
more massive than V2247~Oph), the distribution of rotation periods is bimodal, with 
a main peak centred at about 7.5~d and an overall average of about 6.7~d in the 
Orion Nebula Cluster (ONC);  for stars redder than this threshold, the period 
distribution is unimodal with an average of about 3.3~d in the ONC.  
With a rotation period of 3.5~d, V2247~Oph belongs to the second group while the 
prototypical cTTS BP~Tau (whose $R_{\rm c}-I_{\rm c}$ and rotation period are 
respectively equal to 0.9 and 7.6~d) belongs to the more slowly-rotating 
population\footnote{The exact stellar mass at which this limit occurs 
somewhat depends on the accuracy of both atmospheric and evolutionary models;  
while studies suggest it is located at about 0.25~\msun\ \citep{Lamm05}, our 
mass estimate of V2247~Oph would place it at about 0.35~\msun.}.   

From the photometric period and projected equatorial rotation velocity \citep[equal 
to $20.5\pm0.5$~\kms, see below, in good agreement with the lower accuracy estimate 
of][]{Bouvier86}, we infer that $\rstar \sin i \simeq 1.4$~\rsun, where $i$ denotes 
the inclination of the rotation axis to the line-of-sight.  We obtain that 
$i=45\pm10\degr$;  our imaging study further confirms this intermediate value of the 
inclination angle (see Sec.~\ref{sec:mod}).   

Optical veiling, presumably caused by accretion and potentially weakening the 
strength of the photospheric spectrum, is rather common for cTTSs and often generates 
large-scale and apparently random variations in the equivalent width of photospheric 
lines.  For V2247~Oph however, we see only small changes (of order a few percent) 
in the equivalent widths of LSD Stokes $I$ profiles, suggesting that veiling is minimal;  
we take this as direct evidence that accretion was weak on V2247~Oph during our 
observations.  The spectrum of V2247~Oph also exhibits all the usual accretion 
proxies, namely \hal, \hbe, the \hei\ $D_3$ line and the emission cores of the \caii\ 
infrared triplet (IRT) lines.  From their equivalent widths and the corresponding line 
fluxes (see Sec.~\ref{sec:acc}), and using empirical correlations from the published 
literature \citep[e.g.,][]{Mohanty05, Fang09}, we can derive a rough estimate of the 
logarithmic mass accretion rate \Mdot\ of V2247~Oph (in \mspy) at the time of our run 
of about --9.8;  using the same correlations, we find that \Mdot\ can reach 
$10^{-9}$~\mspy\ during stronger accretion episodes \citep[such as that observed 
by, e.g.,][]{Littlefair04}.  As expected, this is slightly below the typical values 
expected for low-mass cTTSs \citep[e.g.,][]{Mohanty05, Fang09}.  

\begin{table}
\caption[]{Journal of observations.  Columns 1--4 respectively list the UT date, 
the heliocentric Julian date and UT time (both at mid-exposure), 
and the peak signal to noise ratio (per 2.6~\kms\ velocity bin) of each observation 
(i.e., each sequence of $4\times879$~s subexposures).  
Column 5 lists the rms noise level (relative to the unpolarized continuum level 
$I_{\rm c}$ and per 1.8~\kms\ velocity bin) in the circular polarization profile 
produced by Least-Squares Deconvolution (LSD), while column~6 indicates the 
rotational cycle associated with each exposure (using the 
ephemeris given by Eq.~\ref{eq:eph}).  }   
\begin{tabular}{cccccc}
\hline
Date & HJD          & UT      &  \sn\  & $\sigma_{\rm LSD}$ & Cycle \\
(2008) & (2,454,000+) & (h:m:s) &      &   (\ptt)  &  \\
\hline
Jul 19 & 666.78070 & 06:40:26 & 110 & 9.1 & 0.937 \\
Jul 20 & 667.76238 & 06:14:11 & 120 & 8.5 & 1.218 \\
Jul 21 & 668.77056 & 06:26:04 & 130 & 7.5 & 1.506 \\
Jul 22 & 669.76031 & 06:11:25 & 140 & 6.9 & 1.789 \\
Jul 22 & 669.80591 & 07:17:06 & 140 & 6.9 & 1.802 \\
Jul 24 & 671.75990 & 06:11:05 & 130 & 7.7 & 2.360 \\
Jul 25 & 672.77281 & 06:29:47 & 130 & 7.5 & 2.649 \\
Jul 26 & 673.75570 & 06:05:16 & 150 & 6.3 & 2.930 \\
Jul 27 & 674.75594 & 06:05:44 & 140 & 6.8 & 3.216 \\
Jul 27 & 674.88360 & 09:09:35 & 130 & 7.8 & 3.252 \\
Jul 28 & 675.75706 & 06:07:28 & 60 & 18.7 & 3.502 \\
Jul 28 & 675.84904 & 08:19:56 & 50 & 21.6 & 3.528 \\
Jul 29 & 676.75564 & 06:05:32 & 120 & 8.0 & 3.787 \\
Jul 30 & 677.75556 & 06:05:33 & 140 & 7.0 & 4.073 \\
Jul 30 & 677.87649 & 08:59:43 & 120 & 8.5 & 4.108 \\
\hline
\end{tabular}
\label{tab:log}
\end{table}

\section{Observations}
\label{sec:obs}

Spectropolarimetric observations of V2247~Oph were collected in 2008 July 
using ESPaDOnS on the CFHT in Hawaii.  
ESPaDOnS collects stellar spectra spanning the whole optical domain (from 370 
to 1,000~nm) at a resolving power of 65,000 (i.e., 4.6~\kms), in either circular or 
linear polarisation \citep{Donati03}.  A total of 15 circular
polarisation spectra were collected over a period of 12~d;  each polarisation 
spectrum consists of 4 individual subexposures lasting each 879~s and taken in 
different polarimeter configurations to remove all spurious polarisation signatures 
at first order.

Raw frames are processed with {\sc Libre~ESpRIT}, a fully automatic reduction
package/pipeline available at CFHT.  It automatically performs optimal extraction 
of ESPaDOnS unpolarised (Stokes $I$) and circularly polarised (Stokes $V$) spectra 
following the procedure described in \citet{Donati97b}.
The velocity step corresponding to CCD pixels is about 2.6~\kms;  however, thanks
to the fact that the spectrograph slit is tilted with respect to the CCD lines,
spectra corresponding to different CCD columns across each order feature a
different pixel sampling.  {\sc Libre~ESpRIT} uses this opportunity to carry out
optimal extraction of each spectrum on a sampling grid denser than the original
CCD sampling, with a spectral velocity step set to about 0.7 CCD pixel
(i.e.\ 1.8~\kms).
All spectra are automatically corrected of spectral shifts resulting from
instrumental effects (e.g., mechanical flexures, temperature or pressure variations) 
using telluric lines as a reference.  Though not perfect, this procedure provides 
spectra with a relative radial velocity (RV) precision of better than 0.030~\kms\
\citep[e.g.,][]{Donati08d}.

The peak signal-to-noise ratios (per 2.6~\kms\ velocity bin) achieved on the
collected spectra (i.e., the sequence of 4 subexposures) range between 110 and
150, with 2 lower-quality spectra taken in poor weather conditions.  
Rotational cycles $E$ are computed from heliocentric Julian dates according 
to the ephemeris:  
\begin{equation}
\mbox{HJD} = 2454663.5 + 3.5 E. 
\label{eq:eph}
\end{equation}
The full journal of observations is presented in Table~\ref{tab:log}.

\begin{figure}
\center{\includegraphics[scale=0.35,angle=-90]{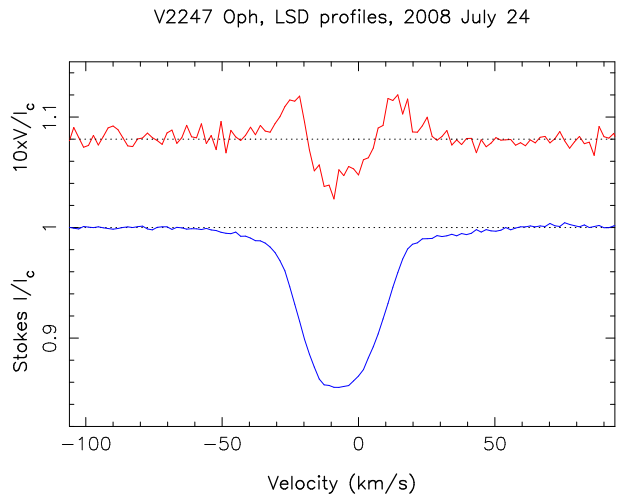}}
\caption[]{LSD circularly-polarized (Stokes $V$) and unpolarized (Stokes $I$) 
profiles of V2247~Oph (top/red, bottom/blue curves respectively) on 2008 July 24 (cycle 2.360).
The mean polarization profile is expanded by a factor of 10 and shifted upwards
by 1.08 for display purposes.  }
\label{fig:lsd}
\end{figure}

Least-Squares Deconvolution (LSD; \citealt{Donati97b}) was applied to all
observations.   The line list we employed for LSD is computed from an {\sc
Atlas9} LTE model atmosphere \citep{Kurucz93} and corresponds to a M2 
spectral type ($\teff=3,500$~K and  $\logg=3.5$) appropriate for V2247~Oph.  
We selected only moderate to strong atomic spectral lines whose synthetic 
profiles had line-to-continuum core depressions larger than 40\% neglecting 
all non-thermal broadening mechanisms.  We omitted the spectral regions with 
strong lines formed mostly outside the photosphere, such as the Balmer and He 
lines, and the \caii\ H, K and infrared triplet (IRT) lines;  we also 
discarded all spectral windows heavily crowded with telluric lines.  
Molecular lines are not included in the list, their wavelengths, intensities 
and Zeeman sensitivities being not empirically computable with enough accuracy; 
we assume that their indirect contribution to LSD profiles is not concentrated at 
discrete velocities but statistically distributed over a wide domain, 
thus inducing no significant change nor variability in the shape 
of Stokes $I$ and $V$ LSD profiles.  
Altogether, about 4,800 spectral features are used in this process, most of
them from \fei.  Expressed in units of the unpolarised continuum level 
$I_{\rm c}$, the average noise levels of the  resulting LSD signatures are 
of order 8$\times10^{-4}$  per 1.8~\kms\ velocity bin, except for the 2 
spectra taken on Jul~28 whose noise is about twice to thrice larger.  

In most spectra, Stokes $I$ LSD profiles show clear distortions tracing the 
presence of cool spots at the surface of V2247~Oph;  in addition, Stokes $V$ 
LSD profiles show Zeeman signatures with peak-to-peak average amplitudes 
of about 0.5\% of the unpolarised continuum (i.e., typically 6 times larger 
than the mean noise level).  An example of LSD profiles is shown in 
Fig.~\ref{fig:lsd}.  We note that Stokes $I$ and $V$ spectra recorded almost 
exactly 7~d apart from each other (thus corresponding to very similar rotation 
phases but different cycles, e.g., cycles 1.218 and 3.216, or cycles 1.789 and 
3.787) show significant differences, larger than those usually attributable 
to small phase shifts (e.g., between cycles 3.216 and 3.252, or cycles 1.789 
and 1.802).  We speculate that this is likely evidence for strong enough 
differential rotation at the surface of V2247~Oph to change the brightness 
and magnetic spot patterns at the surface of the star in as little as 7~d;  
this option is investigated in Sec.~\ref{sec:mod}.

\section{Brightness and magnetic imaging} 
\label{sec:mod}

\subsection{Imaging method}

To model the cool spots and magnetic topology at the surface of V2247~Oph, we use the new 
imaging code of \citet{Donati06b} where the field is described through spherical-harmonics 
expansions and the brightness distribution as a series of independent pixels.  
In its newest implementation, our code can also simultaneously recover the distribution 
of accretion spots as derived from the profiles of emission lines formed at the base of 
accretion funnels (e.g., the narrow emission components of the IRT lines).  Since 
V2247~Oph was experiencing only weak accretion at the time of our observations, we 
keep this part of the modelling separate and describe it in the next section.  

Our imaging code uses the principles of maximum entropy to retrieve the simplest magnetic 
image and brightness distribution compatible with the series of rotationally modulated Stokes 
$I$ and $V$ LSD profiles.  
More specifically, the field is divided into its radial-poloidal, non-radial-poloidal and
toroidal components, each of them described as a spherical-harmonics expansion 
\citep[e.g.,][]{Donati06b};  given
the different rotational modulation of Zeeman signatures that poloidal and toroidal fields
generate, our imaging code appears particularly useful and efficient at producing
dynamo-relevant diagnostics about the large-scale magnetic topologies at the surface of
late-type stars.  
The brightness distribution is simply modeled as a set of independent pixels describing 
the contrast with respect to the quiet photosphere;  for this particular analysis, we 
only allow the code to recover cool spots (i.e., darker than the photosphere). 

The reconstruction process is iterative and proceeds by comparing at each step the synthetic
profiles corresponding to the current image with those of the observed data set.  To compute
the synthetic Stokes $I$ and $V$ profiles, we divide the surface of the star into small grid 
cells (typically a few thousands), work out the specific contribution of each grid cell to the 
Stokes profiles (given the brightness, magnetic field strength and orientation within each 
grid cell, as well as the cell RV, location and projected area) and finally sum 
up contributions of all cells.  Since the problem is partly ill-posed, we stabilise the 
inversion process by using an entropy criterion (applied to the spherical harmonics 
coefficients and to the brightness image pixels) aimed at selecting the image with minimum 
information among all those compatible with the data.  

\begin{figure*}
\center{\hbox{\includegraphics[scale=0.28,angle=-90]{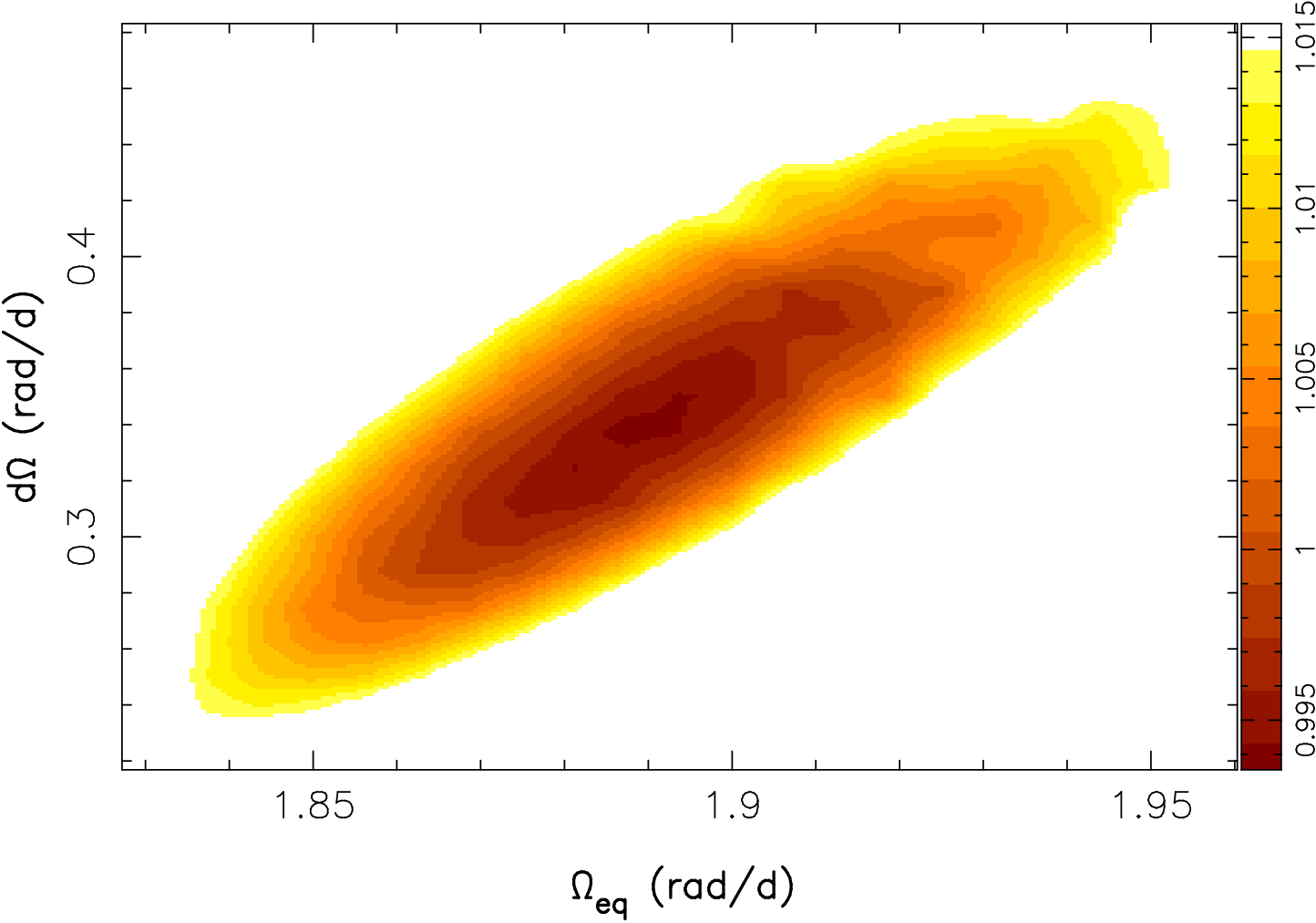}\hspace{3mm} 
\includegraphics[scale=0.28,angle=-90]{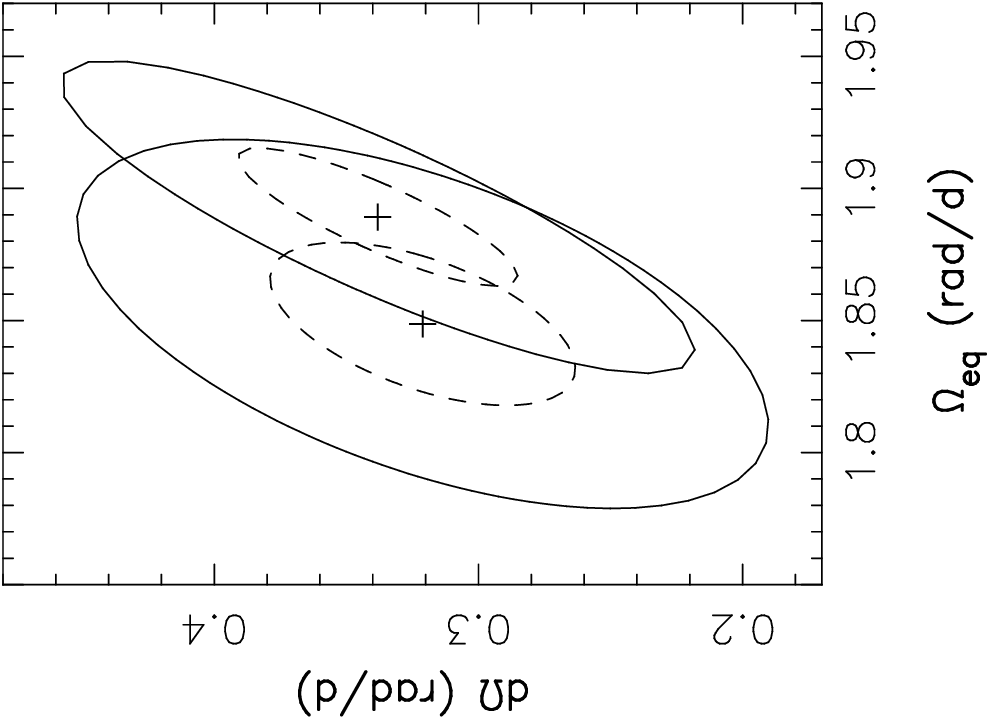}\hspace{3mm} 
\includegraphics[scale=0.28,angle=-90]{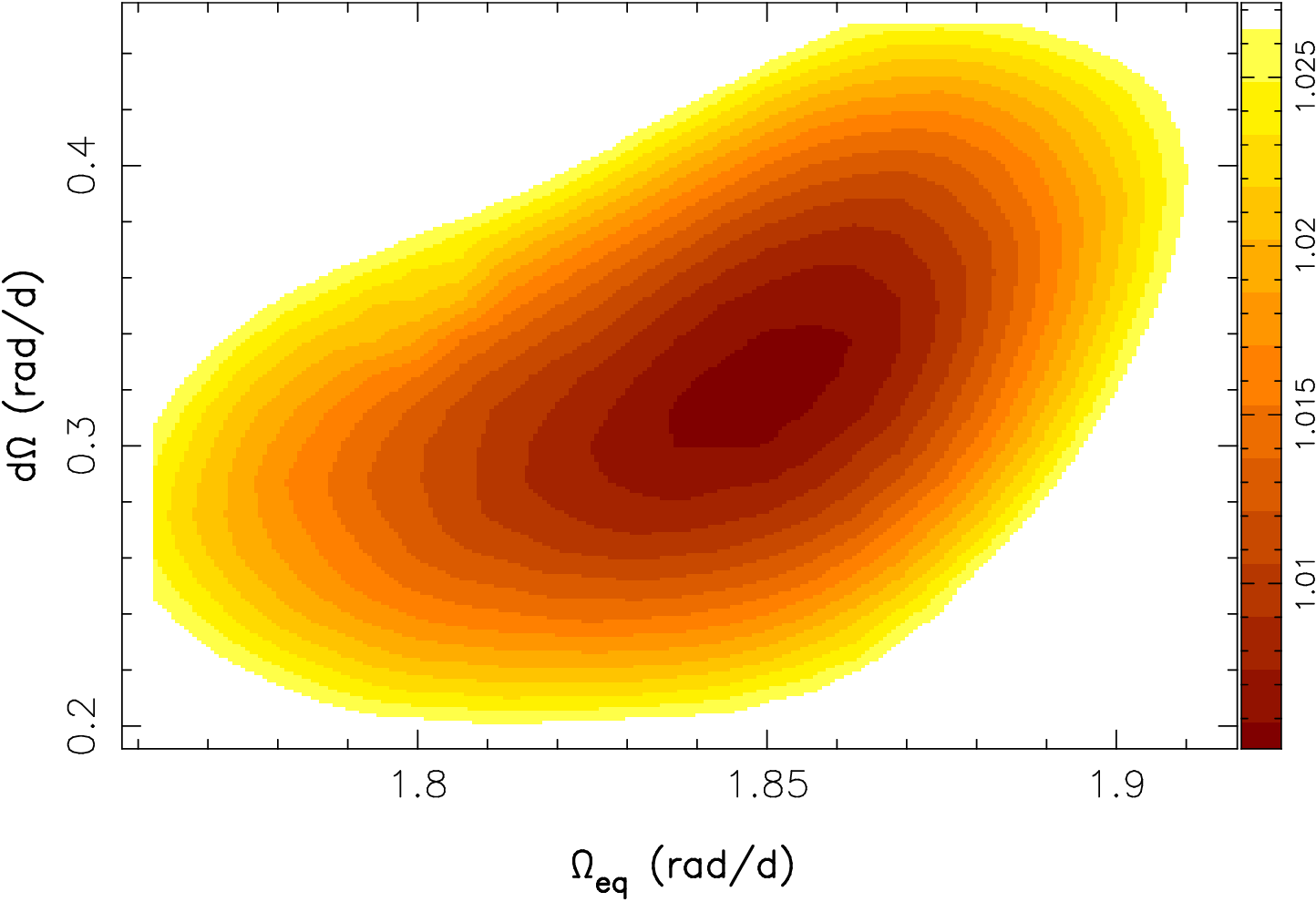}}} 
\caption[]{Variation of \chisqr\ as a function of \omeq\ and \dom, derived from the modelling
of our Stokes $I$ (left) and Stokes $V$ (right) data at constant information content;  
the middle plot shows the corresponding 
confidence ellipses (with 1$\sigma$ and 3$\sigma$ ellipses shown in full and dashed lines).  
For the left and right plots, the outer contour corresponds to a 2.2\% increase in \chisqr\ 
corresponding to a 3$\sigma$ ellipse for both parameters as a pair. } 
\label{fig:drot}
\end{figure*}

To describe the local profiles, we use Unno-Rachkovsky's equations \citep[e.g.,][]{Landi04} 
known to provide a fairly accurate description of Stokes $I$ and $V$ profiles in the presence 
of magnetic fields.  We set the central wavelength, Doppler width and Land\'e factor of our 
equivalent line to 700~nm, 3.4~\kms\ and 1.2 respectively and adjust the average line 
equivalent width to the observed value.  
We then derive both the rotational broadening \vsini\ and the radial
velocity \vrad\ of V2247~Oph by fitting our series of LSD Stokes $I$ profiles and
selecting the values that minimise the image information content (for a given
quality of the fit to the data).  We find $\vsini=20.5\pm0.5$~\kms\ and
$\vrad=-6.3\pm0.5$~\kms.  We can also obtain a rough estimate of 
the inclination angle by selecting the one producing images with minimum information 
content;  the value we derive (when fitting both Stokes $I$ and $V$ data simultaneously) 
is $45\pm10\degr$, in good agreement with our initial estimate (see Sec.~\ref{sec:v22}).  

As noted above, the LSD profiles (and in particular their apparent deviation from strict 
rotational modulation) suggest that the brightness and magnetic distributions may 
have changed significantly over the course of our run.  By comparison with other active 
cool stars, it is unlikely that the field would vary significantly over a period of one 
week;  the probable cause for this change is thus surface differential rotation.  
This conclusion is independently confirmed by the significant 
changes in photometric period reported for V2247~Oph over the last 2 decades \citep{Grankin08}.  
To incorporate differential rotation into the modelling, we proceed as in \citet{Donati03b} and
\citet{Morin08a}, i.e., by assuming that the rotation rate at the surface of the star is 
varying with latitude $\theta$ as  $\omeq - \dom \sin^2 \theta$ where \omeq\ is the rotation 
rate at the equator \dom\ the difference in rotation rate between the equator and the pole.  
When computing the synthetic profiles, we use this law to work out the longitude shift of each 
cell at each observing epoch with respect to its location at the median observing epoch at 
which the field is reconstructed (at rotation cycle 2 in the present case and in the ephemeris 
of Eq.~\ref{eq:eph}, i.e., at HJD~2454670.5);  this way, we can correctly evaluate the true 
spectral contributions of all cells at all epochs.

For each pair of \omeq\ and \dom\ values within a range of acceptable values, we then derive,
from the complete data set, the corresponding magnetic topology (at a given information content)
and the associated reduced chi-squared \chisqr\ at which modelled spectra fit observations.
By fitting a paraboloid to the \chisqr\ surface derived in this process \citep{Donati03b}, we can
easily infer the magnetic topology that yields the best fit to the data along with the corresponding
differential rotation parameters and error bars.  This process has proved reliable for 
estimating surface differential rotation on magnetic stars \citep[e.g.,][]{Donati03b}.

\begin{figure*}
\center{\includegraphics[scale=0.7]{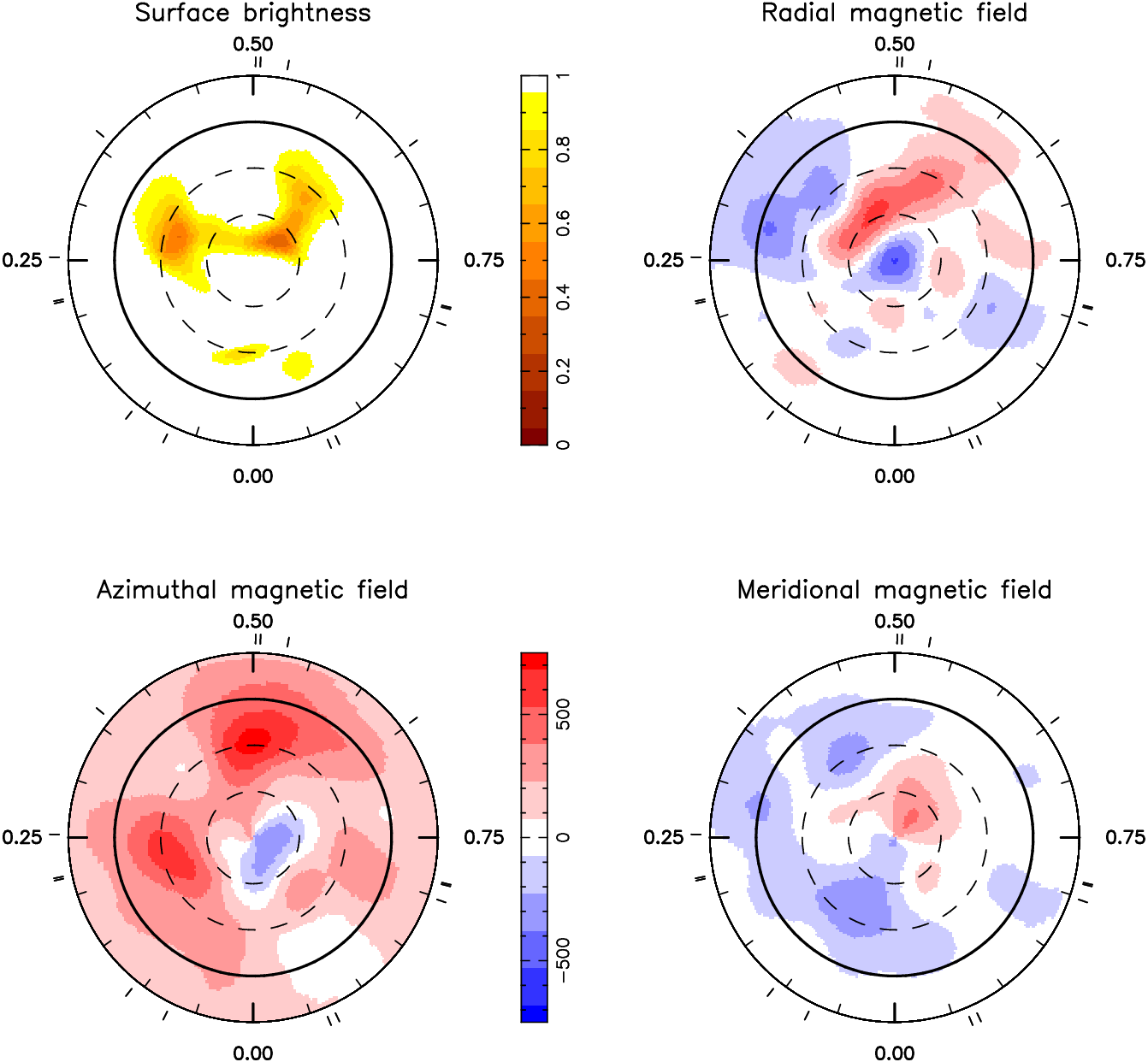}}
\caption[]{Surface brightness (top left) and magnetic topology of V2247~Oph in 2008 
July (HJD~2454670.5), simultaneously derived from the Stokes $I$ \& $V$ data shown in 
Fig.~\ref{fig:fit}.  
The radial, azimuthal and meridional components of the field are displayed from top right 
to bottom left (with magnetic flux values labelled in G).
The star is shown in flattened polar projection down to latitudes
of $-30\degr$, with the equator depicted as a bold circle and
parallels as dashed circles.  Radial ticks around each plot indicate
phases of observations. } 
\label{fig:map}
\end{figure*}

\subsection{Results}

The first result is that both Stokes $I$ and $V$ LSD profiles confirm that differential 
rotation is indeed significantly shearing the photosphere of V2247~Oph.  In both cases, 
the maps describing how \chisqr\ varies as a function of \omeq\ and \dom\ (at constant 
information content) show a clear paraboloid and a well defined minimum (see 
Fig.~\ref{fig:drot}).  
The differential rotation parameters producing an optimal fit to the data 
are respectively equal to $\omeq=1.89\pm0.02$~\rpd\ and $\dom=0.34\pm0.04$~\rpd\ for 
Stokes $I$ data and $\omeq=1.85\pm0.03$~\rpd\ and $\dom=0.32\pm0.05$~\rpd\ for Stokes 
$V$ data.  Both results are compatible within about one $\sigma$ of their weighted 
average ($\omeq=1.87$~\rpd, $\dom=0.33$~\rpd, see middle panel of Fig.~\ref{fig:drot}) 
that we take as reference hereafter.  
The corresponding rotation periods for the equator and pole are 3.36 
and 4.08~d respectively, bracketing the reported photometric periods ranging 
from 3.4 to 3.6~d \citep{Grankin08};  in this context, the period with which we phased 
our data (3.5~d) corresponds to a latitude of about 30\degr.  

We carried out several tests to confirm that differential rotation is indeed detected.  
For instance, trying to fit Stokes $V$ LSD profiles to a unit reduced chi-squared \chisqr\ 
assuming solid body rotation is hardly possible.  The smallest \chisqr\ we get is about 13\% higher 
(i.e., \chisq\ larger by 70 for 540 data points) than the minimum obtained when assuming differential 
rotation at equal information content, and the corresponding rotation period (3.75~d) is significantly 
larger than the average photometric period;  \chisqr\ even gets up to 25\% higher (i.e., 
\chisq\ larger by 135) than the minimum when imposing solid body rotation at a period of 
3.5~d.  We also investigated whether the observed differential rotation could actually be 
a spurious signal resulting from intrinsic variability (other than differential rotation);  
note that this option is unlikely as there is little chance for intrinsic variability 
to mimic a fast equator and a slow pole in the same way in 2 different and fully-independent 
data sets.  We nevertheless checked it by splitting our data into 2 subsets and 
reconstructing individual images for each subset (assuming solid body rotation);  although this 
method has limited diagnostic power (the fewer the spectra in one subset the less accurate the 
recovered image), it is nevertheless clear that both images include similar features at 
slightly different positions and different contrasts.  
We therefore believe that our detection is real.  

\begin{figure*}
\center{\hbox{\includegraphics[scale=0.6,angle=-90]{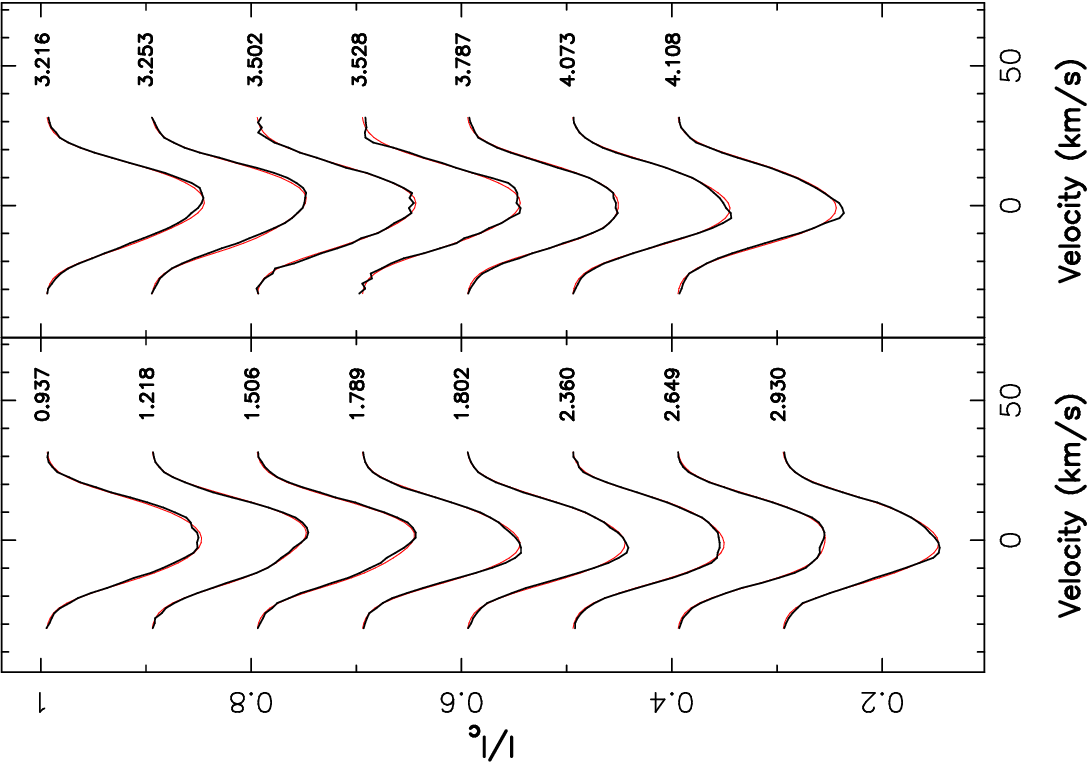}\hspace{10mm}
\includegraphics[scale=0.6,angle=-90]{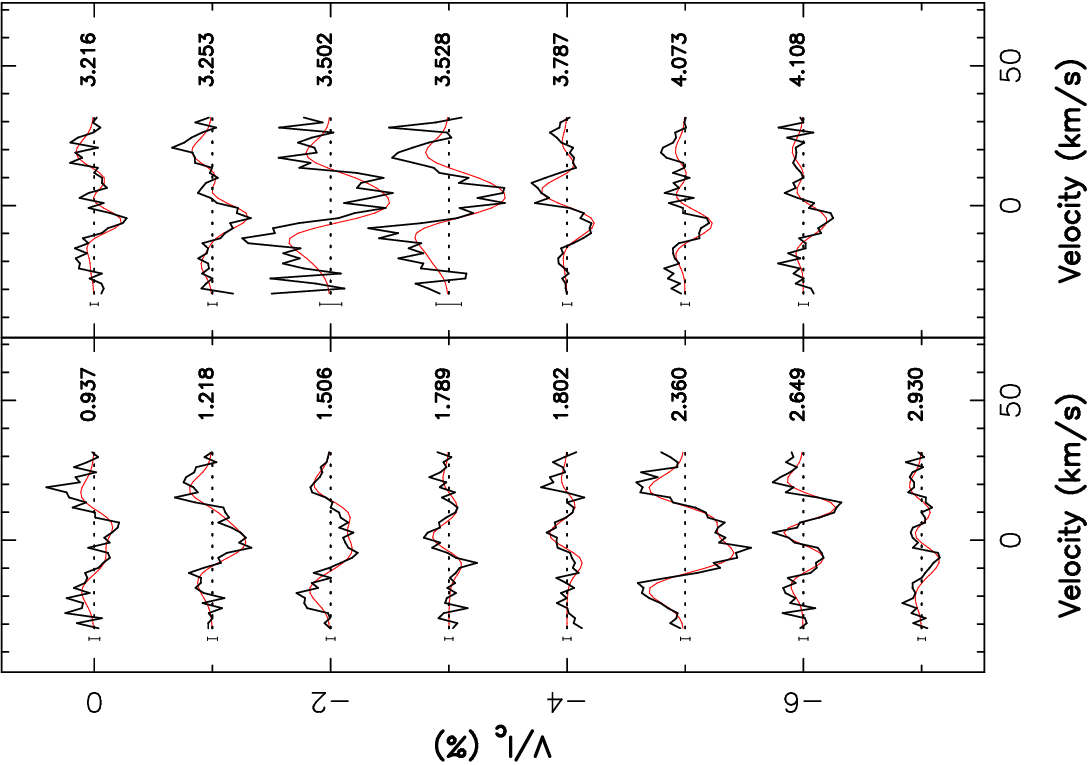}}} 
\caption[]{Maximum-entropy fit (thin red line) to the observed Stokes $I$ (left) 
and Stokes $V$ (right) LSD photospheric profiles (thick black line) of V2247~Oph.  The rotation 
cycle of each observation (as listed in Table~1) and 1$\sigma$ error bars (for Stokes $V$ profiles) 
are also shown next to each profile.  }
\label{fig:fit}
\end{figure*}

The surface brightness and magnetic map we derive (see Fig.~\ref{fig:map}) corresponds to 
a unit \chisqr\ fit to the data (see Fig.~\ref{fig:fit}), the initial \chisqr\ 
(corresponding to an unspotted and non-magnetic star) being about 
3.4 for 1080 data points in total.  The brightness map is fairly simple and includes 2 
main features at phases 0.30 and 0.60 with latitudes ranging from 30 to 70\degr.  In 
particular, V2247~Oph does not feature a polar cap like the conspicuous ones found, e.g., 
on the higher mass cTTS V2129~Oph \citep{Donati07} and BP~Tau \citep{Donati08b}.  
Spots large and dark enough to affect significantly the shape of Stokes $I$ LSD profiles 
do not cover more than 3.5\% of the stellar surface, at least at the time of our 
observations.  Given the significant rotational broadening in the spectrum of V2247~Oph, 
this result is well constrained and reliable and cannot be attributed to potential 
defects of surface imaging.  

The large-scale magnetic field of V2247~Oph that we recover is also fairly different from 
those of V2129~Oph and BP~Tau \citep{Donati07, Donati08b};  not only is the detected magnetic 
flux much weaker (about 300~G in average) but its topology is also fairly different, with a 
strong toroidal component (totalling about 60\% of the magnetic energy, as obvious from 
Fig.~\ref{fig:map}) and a predominantly non-axisymmetric poloidal component (less than 
30\% of the poloidal field energy concentrating in modes with $m<\ell/2$).  Moreover, 
the poloidal field is complex with about half of the poloidal field energy concentrating 
in modes with $\ell>2$;  this is fairly obvious from the radial field component, featuring 
nearby regions of opposite polarities (see Fig.~\ref{fig:map}).  

To investigate further the large-scale magnetic topology of V2247~Oph, we forced the code 
to recover fields that are either mainly symmetric or antisymmetric with respect to the 
centre of the star;  this is achieved by favouring spherical harmonics modes of even or 
odd $\ell$ values respectively.  We find that a symmetric field configuration is marginally 
more likely;  in both cases though, the data can easily be fitted down to noise level and 
the recovered topologies in the visible hemisphere are very similar.  
The dipole component of the poloidal field is weak and strongly tilted with respect to 
the rotation axis;  its polar strength ranges from 80 to 100~G and concentrates about 
25--30\% of the poloidal field energy (depending on the assumed symmetry properties).  

Note that the intrinsic weakness of the dipole component on V2247~Oph is a 
reliable and well constrained result;  in particular, it cannot be attributed to 
observational problems, e.g., to magnetic flux hiding in large high-contrast dark spots 
(absent on V2247~Oph at the time of our observations) or to flux cancellation from 
regions of opposite polarities (easily detected in the radial field component).

\section{Activity \& accretion}
\label{sec:acc}

As usual for cTTSs, V2247~Oph exhibits in its optical spectrum a number of emission 
lines known to be proxies of activity and/or accretion -- and in particular, the Balmer 
lines \hal\ and \hbe, the \hei\ $D_3$ line and the 849.8~nm line of the \caii\ IRT.  
The average equivalent widths that we measure in these proxies are equal to 137, 140, 
8.5 and 12~\kms\ respectively, or equivalently 0.30, 0.23, 0.017 and 0.035~nm;  
they correspond to logarithmic line fluxes of 6.2, 5.8, 4.9 and 5.4 (with fluxes in 
\epspcs), or equivalently --4.6, --5.0, --5.9 and --5.4 (with fluxes in \lsun).  
From these, we derive an average logarithmic accretion luminosity (in \lsun) of $-3.2\pm0.3$  
and a logarithmic mass accretion rate (in \mspy) of $-9.8\pm0.3$ \citep{Fang09}.  

In this paper, we concentrate mostly on 3 proxies, \hal, \hbe\ and the LSD average of 
all 3 \caii\ IRT lines (averaging out to an equivalent width of about 14~\kms\ or 
0.04~nm).  The \hei\ $D_3$ line is dim and noisy and hardly usable for the present 
analysis.  The complete series of profiles for all 
three proxies are shown in Fig.~\ref{fig:acc};  small Stokes $V$ signatures are 
also detected in conjunction with the IRT lines.  

\begin{figure*}
\center{\hbox{\hspace{-6mm}\includegraphics[scale=0.47,angle=-90]{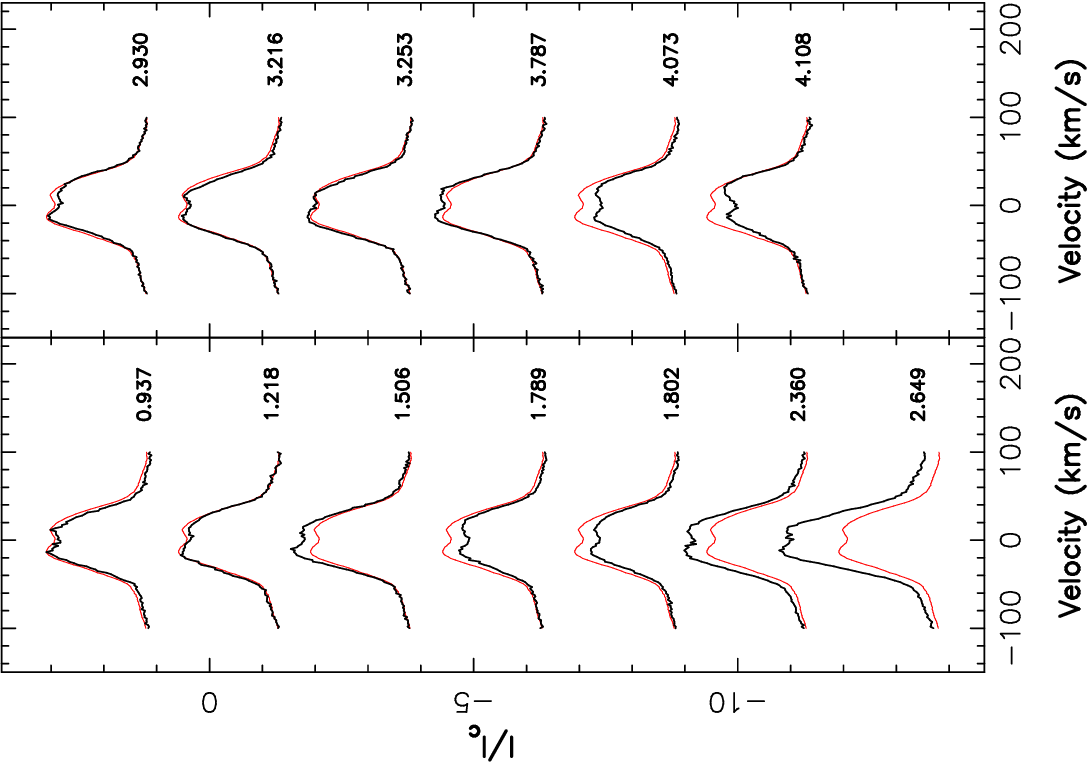}\hspace{2mm}
\includegraphics[scale=0.47,angle=-90]{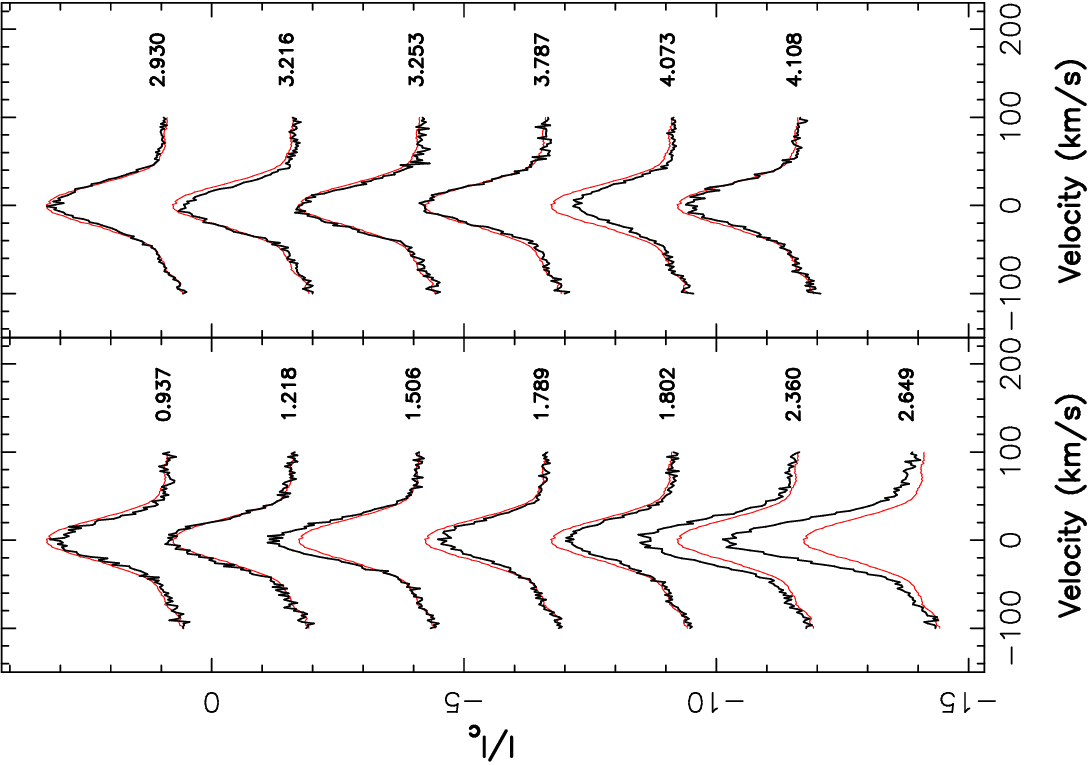}\hspace{2mm} 
\includegraphics[scale=0.47,angle=-90]{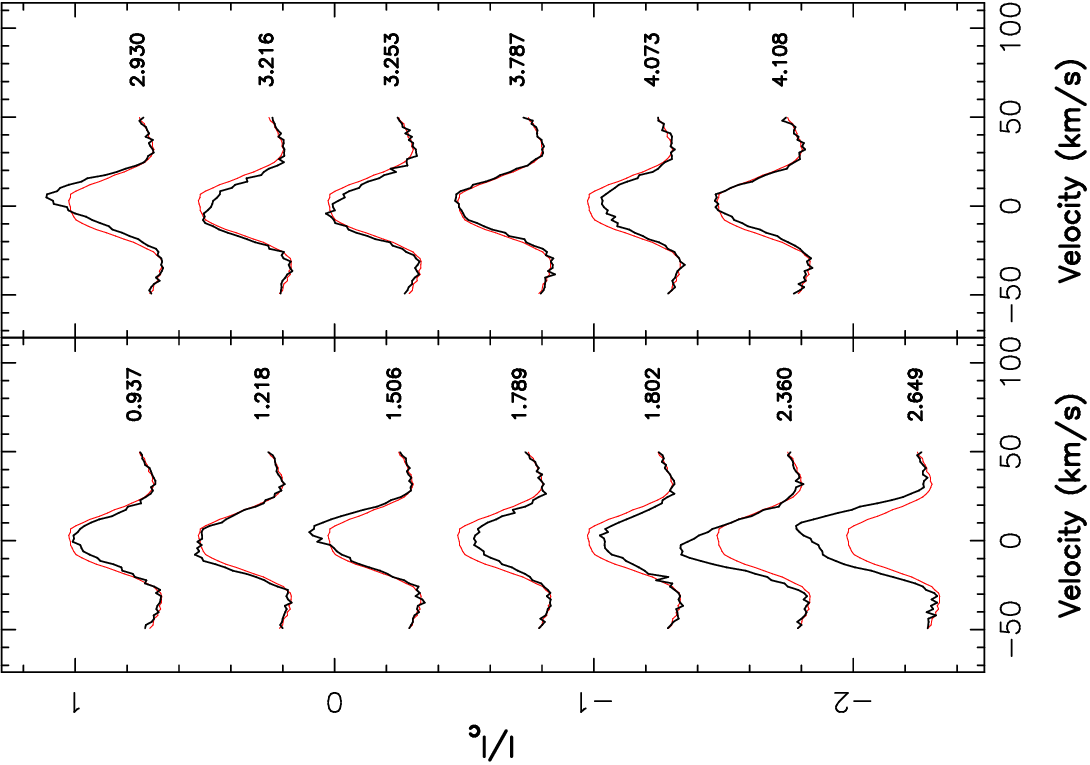}}} 
\caption[]{Variation of \hal\ (left), \hbe\ (middle) and IRT (right) profiles throughout 
our observing run.  In each case, the average profile over the run is shown in red.  
The rotation cycle of each observation (as listed in Table~1) are also shown next to 
each profile.  }
\label{fig:acc}
\end{figure*}

As obvious from Fig.~\ref{fig:acc}, the accretion proxies feature a significant 
level of intrinsic variability.  For instance, the amount of emission at cycle 
1.789 is clearly different from that 2 full rotation cycles later (at cycle  
3.787) in all 3 proxies;  the emission peak detected at cycles 2.360 and 2.649 
also seems fairly difficult to reconcile with emission fluxes at nearby phases 
in both the previous and following rotation cycles.  
To separate intrinsic variability from rotational modulation (generating mostly 
low frequency periodic variations), we propose the following procedure.  We 
start by computing the equivalent widths of all 3 proxies as a function of 
rotation cycle (excluding the low quality data collected on 28~July) and fit 
them with a low-frequency periodic wave -- hopefully retaining rotational 
modulation only and filtering out intrinsic variability;  we then simply 
rescale all profiles to their fitted equivalent widths.  While obviously no 
more than approximate, this procedure has the advantage of being very 
straightforward yet reasonably efficient\footnote{Our scaling procedure has 
little impact on the reconstructed maps and mostly aims at filtering out 
the discrepancies between the synthetic and observed profiles that the model 
was not designed to reproduce;  in other words, it essentially makes the code 
convergence quicker and more reliable by emphasising the relevant contrasts 
in the \chisq\ landscape.  }.  

The periodic wave we use to fit the equivalent widths is of the form:  
\begin{equation}
\Sigma_{i=1}^2 \ a_i \cos(2 i \pi E/p) + b_i \sin(2 i \pi E/p) + c 
\label{eq:wav}
\end{equation}
where $E$ denotes the rotation cycle of each data point (as in Eq.~\ref{eq:eph}) 
and $p$, $a_i$, $b_i$ and $c$ are the 6 free parameters of our model wave.  
We find that for the 3 proxies, the fit residuals show a very clear minimum for 
$p\simeq1.08$, i.e., for a period of about 3.8~d.  Assuming that the differential 
rotation law derived in Sec.~\ref{sec:mod} also applies in the chromosphere 
(where accretion proxies supposedly form), it suggests that active and/or 
accreting regions mainly concentrate at intermediate latitudes of about 50\degr.  
An example of the modelled rotational modulation in the particular case of 
the IRT is shown in Fig.~\ref{fig:ew}.  

\begin{figure}
\center{\includegraphics[scale=0.35,angle=-90]{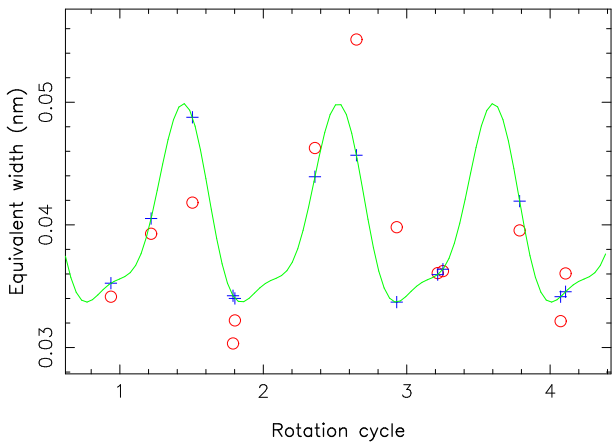}}
\caption[]{Measured (red open circles) and fitted (blue crosses) equivalent widths of 
the \caii\ IRT LSD profiles throughout our observing run.  The model wave (green line) 
providing the best fit to the data features a period of about 3.8~d.  }  
\label{fig:ew}
\end{figure}

By rescaling all IRT profiles to their fitted equivalent widths, we obtain 
a data set that can be added to that from photospheric lines (see 
Sec.~\ref{sec:mod}) and further constrain the modelling of V2247~Oph.  
In that respect, we proceed as in previous work \citep{Donati07, Donati08b}
and assume that IRT emission includes both a quiet chromospheric contribution 
from the whole stellar surface as well as a stronger but more localised 
contribution from accretion regions.  In previous work, we assumed (as 
a first guess) that accretion spots coincide with dark photospheric spots;  
we no longer use this approximation here and recover 2 completely 
independent distributions (in addition to the magnetic field), one giving 
the photospheric surface brightness (as in Sec.~\ref{sec:mod}) and another 
one describing the amount of excess emission from accretion spots.  The local 
emission line (modelled again using Unno-Rachkovsky's equations) features  
a central wavelength, Doppler width and Land\'e factor of 850~nm, 7~\kms\ and 
1.0 respectively.  The intrinsic emission within accretion spots is 
(arbitrarily) assumed to be 10 times that within the quiet photosphere;  
the exact value of this ratio has little impact on the result, e.g., with 
smaller ratios producing only more contrasted accretion regions without 
noticeably changing their distribution.    

\begin{figure}
\center{\includegraphics[scale=0.35]{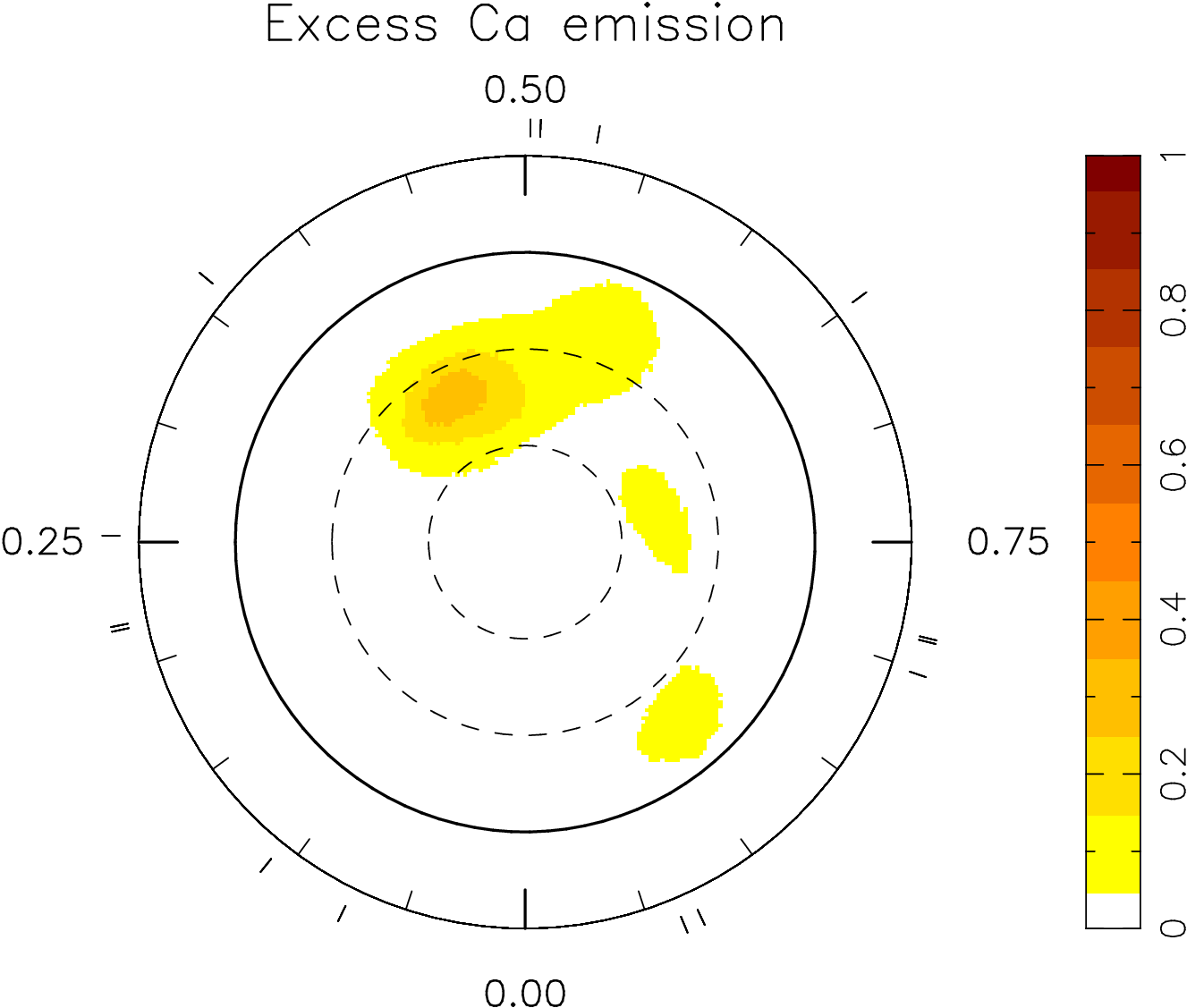}}
\caption[]{Distribution of excess Ca IRT emission at the surface of V2247~Oph.} 
\label{fig:map2}
\end{figure}

\begin{figure*}
\center{\hbox{\includegraphics[scale=0.6,angle=-90]{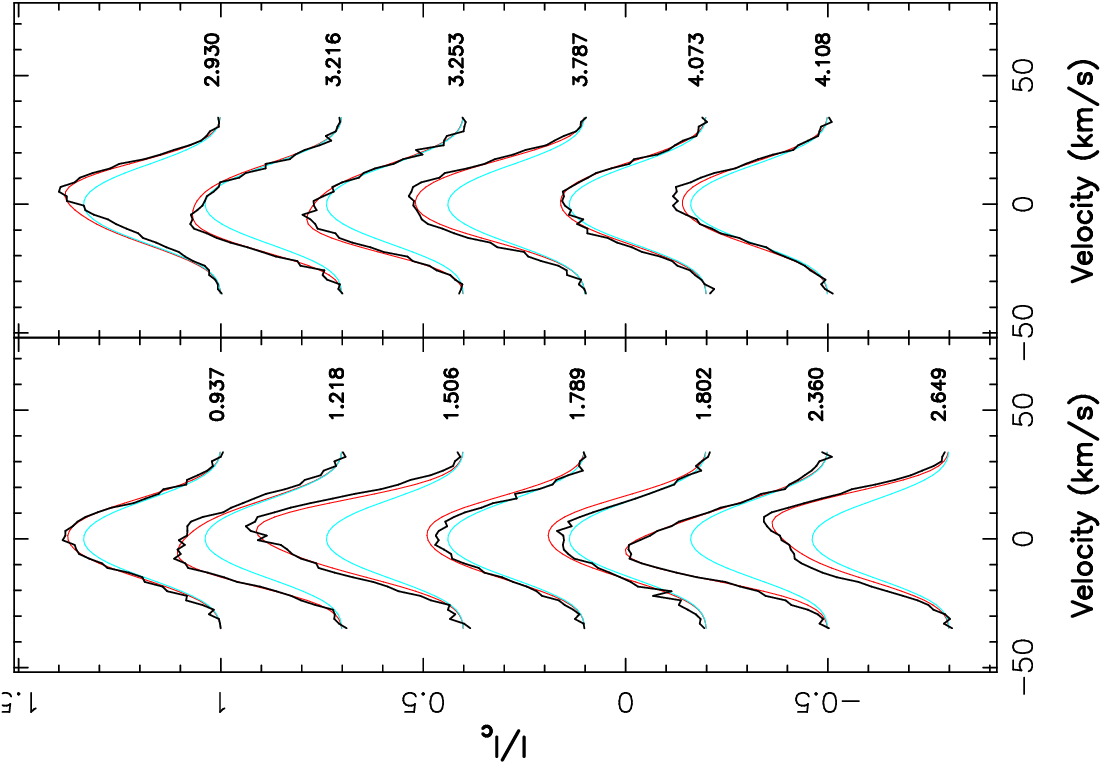}\hspace{10mm}
\includegraphics[scale=0.6,angle=-90]{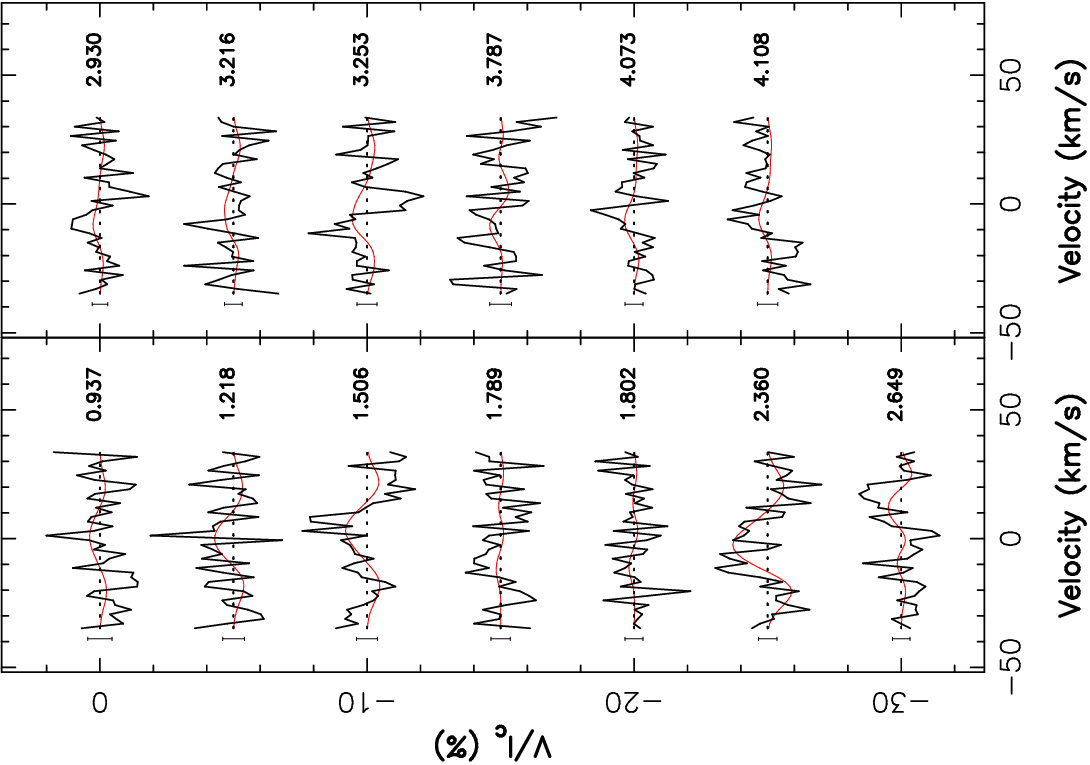}}} 
\caption[]{Maximum-entropy fit (thin red line) to the observed Stokes $I$ (left) 
and Stokes $V$ (right) IRT emission profiles (thick black line) of V2247~Oph;  the 
light-blue curve shows the contribution of the quiet chromosphere.  Note that all 
profiles were rescaled following the procedure described in the text.  
The rotation cycle of each observation and 1$\sigma$ error bars (for Stokes $V$ 
profiles) are also shown next to each profile.  }
\label{fig:fit2}
\end{figure*}

The surface distribution of excess IRT emission we obtain is shown in 
Fig.~\ref{fig:map2} while the corresponding fit to the Stokes $I$ and 
$V$ profiles is shown in Fig.~\ref{fig:fit2}.  We find that excess IRT 
emission is concentrated in one main region located at phase 0.45 and latitude 
45\degr, and covering about 1.5\% of the stellar surface;  unsurprisingly, 
this result is compatible with the observed modulation of the \caii\ IRT 
core emission (peaking at phase 0.45 and fluctuating with a period corresponding 
to intermediate latitudes, see Fig.~\ref{fig:ew}).  
The brightness and magnetic 
maps at photospheric level (as well as the corresponding fits to Stokes $I$ 
and $V$ LSD profiles) are virtually identical to those already shown in 
Sec.~\ref{sec:mod}.  
The shapes of Stokes $I$ IRT profiles 
(and in particular the asymmetries) are well reproduced in our model, 
ensuring at the same time that the corresponding RVs (observed 
to vary by about $\pm2$~\kms\ about the line centre) are also 
grossly fitted (see Fig.~\ref{fig:rv}).  Given how well correlated the 
equivalent width and RV curves are for the IRT line, \hal\ and \hbe, we 
assume that the excess IRT emission spot we detect is also what generates 
the observed excess \hal\ and \hbe\ emission.  

\begin{figure}
\center{\includegraphics[scale=0.35,angle=-90]{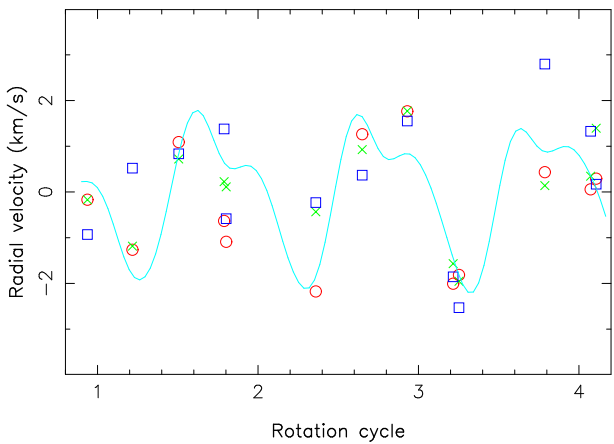}}
\caption[]{Measured RVs of the IRT (red circles), \hal\ (green crosses) and \hbe\ 
(blue squares) profiles throughout our observing run.  The light-blue line shows 
the RV curve that our model predicts.  }  
\label{fig:rv}
\end{figure}

Worth noticing is that the average excess-emission spot 
latitude $l$ that we obtain (about 45\degr) is much less than that one would 
derive from naively assuming that all emission comes from the accretion 
spot and hence that the semi-amplitude of the line RV variations is roughly 
given by $\vsini \cos l$ (which would imply $l\simeq85$\degr).  
The excess emission from accretion spots (generating the RV variations) is 
actually mixed with the (constant and dominant) contribution from the quiet 
chromosphere (the light-blue line in the left panel of Fig.~\ref{fig:fit2}), 
thereby strongly reducing the observed amplitude of the RV 
variations.  The shape of IRT profiles visually demonstrates 
this;  the IRT line is not wiggling back and forth in velocity space as a 
whole (as expected from emission concentrating in a localised accretion spot) 
but features a roughly constant centroid and width at about 10\% height (as 
expected when the whole star significantly contributes to emission 
through a quiet chromosphere).  We thus (unsurprisingly) conclude that, 
fitting emission line profiles (whenever possible) is much more efficient 
and reliable than modelling RV variations only, for mapping accretion spots 
at the surfaces of cTTSs.

\section{Summary and discussion}
\label{sec:dis}

We presented in this paper the first magnetic map for the low-mass cTTS V2247~Oph.  
This is the lowest mass cTTS to be magnetically imaged yet;  despite it being in a low-state 
of accretion (with $\log \Mdot \simeq -9.8$ with \Mdot\ expressed in \mspy), this study brings 
a number of new and unexpected results.  

The brightness distribution we derive at the surface of V2247~Oph (from the 
rotational modulation of Stokes $I$ photospheric LSD profiles) is relatively 
simple, with dark spots covering about 3.5\% of the total surface;  in particular, 
it does not feature a highly-contrasted dark polar spot as seen, e.g., on 
V2129~Oph \citep{Donati07}, BP~Tau \citep{Donati08b} or other cTTSs \citep{Hussain09}.  
It is however roughly similar to what is found on low-mass fully-convective dwarfs 
where large-scale spots are usually low-contrast and cover only a small fraction 
of the stellar surface \citep[e.g.,][]{Morin08a}.  

We also detect clear Zeeman signatures from V2247~Oph, tracing a rather complex 
and moderately strong multipolar large-scale magnetic topology with an average strength 
of 300~G and a mixed amount of poloidal and toroidal field;  the poloidal field 
is mostly non-axisymmetric and features a weak (tilted) dipole component of about 
80--100~G.  Again, this is fairly different from what is seen on intermediate-mass cTTSs 
like BP~Tau, where the field is far simpler, predominantly poloidal, axisymmetric 
with a dipole component exceeding 1~kG \citep{Donati08b}.  
This is also in contrast to low-mass fully-convective mid-M dwarfs 
\citep[e.g.,][]{Morin08b} that almost always exhibit very intense, mainly 
potential large-scale fields with a simple topology roughly aligned with 
the rotation axis.  
New results however indicate that a significant fraction of very-low-mass 
dwarfs (late-M dwarfs with $\mstar<0.2$~\msun, i.e., below about half the 
mass threshold for full convection on the main sequence) are found to host 
moderate and complex non-axisymmetric large-scale magnetic fields (Morin et al 
2009, in preparation), very different from the strong and simple fields of 
mid-M dwarfs.  Although V2247~Oph is significantly more massive (at 0.35~\msun) 
than late-M dwarfs, the difference in evolutionary stage may easily 
compensate for the discrepancy;  the mass of V2247~Oph is indeed also lower 
than half the mass threshold for full convection at an age of about 1~Myr.  
Further observations are of course needed to confirm whether this analogy 
truly holds;  if so, it would bring additional evidence that magnetic 
fields of intermediate- and low-mass cTTSs are similar in nature to those 
of mid-M and late-M dwarfs and are thus produced through dynamo processes 
(rather than being fossil leftovers from an earlier formation stage).  

In addition, we find that the spot distribution and the magnetic field of 
V2247~Oph evolve on a very short timescale, of order one week;  in both 
cases, this evolution is compatible with surface differential rotation shearing 
the photosphere 5 to 6 times faster than on the Sun, with the equator 
lapping the pole by one complete rotation every $19\pm2$~d.  This is 
again fairly unexpected since all fully-convective stars have been reported 
to show little to no differential rotation up to now \citep[e.g.,][]{Donati06a, 
Morin08a, Morin08b}.  Only late F stars with very shallow convective zones are 
yet known to exhibit a similar degree of photospheric shear 
\citep[e.g.,][]{Donati08c}.  Further confirmation from new data sets 
on V2247~Oph (and other similar low-mass cTTSs) are thus needed to validate 
this surprising result.  

The amount of differential rotation we estimate is compatible with the range of 
photometric periods measured on V2247~Oph over the last 2 decades \citep[from 3.4 to 
3.6~d, implying $\dom>0.1$~\rpd,][]{Grankin08}.  Given the differential 
rotation law we derive, these periods suggest that spots preferentially 
cluster at low to intermediate latitudes (15--40\degr) on V2247~Oph;  
this is at least qualitatively compatible with the fact that no 
high-contrast polar spot is present at the surface.  
This is the first time that differential rotation is detected on a 
cTTS;  the recent claim \citep{Herbst06} that the high-mass cTTS HBC~338 
(spectral type G9) is differentially rotating (given the observed 
variations of the photometric period) needs confirmation, cTTSs with massive 
accretion discs being prone to photometric perturbations (including sudden 
changes of the light-curve period) likely caused by accretion 
\citep[e.g.,][]{Simon90, Donati08b} rather than resulting from 
differential rotation.  

Finally, we find that V2247~Oph features a region of excess IRT (and \hal\ and 
\hbe) emission at phase 0.45 and intermediate latitude.  
There is no obvious correlation between the location of this emission region 
and the cool spots mapped from the distorted LSD profiles of photospheric 
lines;  however, we find that it roughly coincides with one of the strongest 
field region detected on V2247~Oph, the positive radial field spot whose 
magnetic flux reaches about 0.5~kG.  We speculate that this spot may be the 
footpoint of an accretion funnel linking the star to its accretion disc 
at the time of our observations.  If confirmed, we expect such accretion 
funnels to be rather short lived, with differential rotation continuously 
distorting the magnetic connections between the star and the disc as well 
as the stellar field itself.  In particular, this effect may be partly 
responsible for the sporadic accretion episodes observed on V2247~Oph, with 
more intense accretion episodes occurring when the magnetic topology linking 
the star to the disc is favouring accretion.  

In addition to the differences in their magnetic topologies,  V2247~Oph and 
BP~Tau show quite distinct rotational properties;  as emphasised in 
Sec.~\ref{sec:v22}, each of them belongs to a distinct population of young 
stars \citep{Lamm05}, V2247~Oph to the low-mass cTTSs with short rotation 
period (of about half a week in average) while BP~Tau to the intermediate-mass 
cTTSs with long rotation periods (of about one week in average).  
These two issues may relate together, i.e., the magnetic topology may have 
an impact on the rotation period through a coupling mechanism such as, 
e.g., the well-known disc-locking picture \citep{Camenzind90, Konigl91}.  
In this model, the rotation period at the surface of the star settles close 
to the Keplerian period of the point in the disc (called the Alfven radius 
and noted \rmag) where the ram pressure of the average accretion flow roughly 
equals the magnetic pressure of the largest-scale (i.e., the dipole component) 
poloidal field \citep[e.g.,][]{Bessolaz08}.  Using Eq.~1 of \citet{Bessolaz08} 
(with $B_*$ set to the average dipole field strength at the rotation equator, 
i.e., about 40~G for V2247~Oph), we estimate $\rmag\simeq3.9$~\rstar;  
we can also work out that the radius at which the Keplerian period matches 
the average rotation period (called the corotation radius and noted \rcor) 
is equal to $\rcor\simeq3.4$~\rstar\ for V2247~Oph.  Both estimates roughly 
agree with each other, suggesting a reasonable agreement with the disc-locking 
model predictions.  

Comparing directly with BP~Tau however \citep[and assuming a logarithmic 
mass accretion rate of --7.5 taken from the literature as in][]{Donati08b}, the 
model predicts that both stars should have similar \rmag\ and thus behave 
similarly with respect to the disc-locking mechanism, the 15-fold weaker 
dipole field of V2247~Oph being almost 
compensated by the 200-fold weaker mass accretion rate;  re-estimating 
$\log \Mdot$ for BP~Tau using the same procedure as for V2247~Oph, we 
actually find a much smaller value, equal to $-8.6\pm0.4$.  
While it is not clear whether this new estimate is more accurate than 
the one used previously, it is at least more consistent with the one 
we derived for V2247~Oph when it comes to comparing both stars;  
in this case, we obtain that $\rmag\simeq7.1$~\rstar\ for BP~Tau 
(taking $B_*\simeq600$~G) in reasonable agreement with 
$\rcor\simeq7.4$~\rstar.  

Obviously, more stars of different masses and rotation periods, 
and with consistent estimates of $B_*$ and \Mdot, 
are needed to assess the validity of the disc-locking picture;  
if confirmed, it may ultimately explain why low-mass cTTSs rotate 
in average faster than their intermediate mass equivalents.
The MaPP program should bring new results along this line in 
the near future.

\section*{Acknowledgements}

We thank the CFHT/QSO staff for their efficiency at collecting data;  we are 
also grateful for valuable comments from an anonymous referee that contributed 
to improve the paper.  
This work was supported by the French ``Agence Nationale pour la Recherche'' (ANR) 
within the ``Magnetic Protostars and Planets'' (MaPP) project.

\bibliography{v2247}

\bibliographystyle{mn2e}

\end{document}